\documentstyle[11pt,epsf]{article}
\input psfig
\def\beq{\begin{eqnarray}}
\def\eeq{\end{eqnarray}}
\def\bea{\begin{eqnarray*}}
\def\eea{\end{eqnarray*}}

\def\centeron#1#2{{\setbox0=\hbox{#1}\setbox1=\hbox{#2}\ifdim
\wd1>\wd0\kern.5\wd1\kern-.5\wd0\fi
\copy0\kern-.5\wd0\kern-.5\wd1\copy1\ifdim\wd0>\wd1
\kern.5\wd0\kern-.5\wd1\fi}}
\def\ltap{\;\centeron{\raise.35ex\hbox{$<$}}{\lower.65ex\hbox{$\sim$}}\;}
\def\gtap{\;\centeron{\raise.35ex\hbox{$>$}}{\lower.65ex\hbox{$\sim$}}\;}

\def\singleandthirdspaced{\baselineskip=\normalbaselineskip\multiply
    \baselineskip by 130\divide\baselineskip by 100}
\def\singlespaced{\baselineskip=\normalbaselineskip}
\def\dslash{\not{\hbox{\kern-2pt $\partial$}}}
\def\Dslash{\not{\hbox{\kern-4pt $D$}}}
\def\Oslash{\not{\hbox{\kern-4pt $O$}}}
\def\Qslash{\not{\hbox{\kern-4pt $Q$}}}
\def\pslash{\not{\hbox{\kern-2.3pt $p$}}}
\def\kslash{\not{\hbox{\kern-2.3pt $k$}}}
\def\lslash{\not{\hbox{\kern-2.3pt $l$}}}
\def\qslash{\not{\hbox{\kern-2.3pt $q$}}}
\def\epsilonslash{\not{\hbox{\kern-2.3pt $\epsilon$}}}
\newcommand{\newc}{\newcommand}
\newc{\qbar}{{\overline q}}
\newc{\Kahler}{K\"ahler }
\newc{\deltaGS}{\delta_{\rm GS}}
\begin{document}
\begin{titlepage}
\begin{flushright}
{\large hep-ph/0106356 \\ SCIPP-01-26\\
}
\end{flushright}
\vskip 1.2cm
\begin{center}
{\LARGE\bf Comments on Non-Commutative Phenomenology}
\vskip 1.4cm
{\large  Alexey Anisimov, Tom Banks, Michael Dine and Michael Graesser}
\\
\vskip 0.4cm
{\it Santa Cruz Institute for Particle Physics,
     Santa Cruz CA 95064  } \\
\vskip 4pt
\vskip 1.5cm
\begin{abstract}
It is natural to ask whether non-commutative geometry plays a role in
four dimensional physics.
By performing explicit computations in various toy models, we show
that quantum effects
lead to violations of Lorentz invariance at the level of
operators of dimension three or four.  The resulting constraints
are very stringent.
\end{abstract}
\end{center}
\vskip 1.0 cm
\end{titlepage}
\setcounter{footnote}{0} \setcounter{page}{2}
\setcounter{section}{0} \setcounter{subsection}{0}
\setcounter{subsubsection}{0}
\singleandthirdspaced
\section{Introduction}
Mathematicians proposed non-commutative geometry and the
corresponding non-commutative field theories as a possible
alternative to conventional field theories\cite{ncreview}.
Physicists, for some time, were skeptical about the possible role
of such theories, given their non-local character. However,
interest increased with the realization that such theories can
appear as the low energy limit of string theory in certain regions
of the moduli  space\cite{sw}.  This has lead a number of authors, to ask
whether such non-commutativity might play a role in physics, and
particularly whether non-commutative field theory might become
manifest at accessible energy 
scales\cite{rizzo,stronglimits,ncqed,hinchliffe,harveyetal,jackiw}.
Unlike the situation for supersymmetry or large dimensions, we do
not currently possess a compelling argument that non-commutativity
should be relevant, or that it should be relevant at some
particular energy scale. Indeed, it is not clear that we can write
down a non-commutative generalization of the standard model in a
simple way. Still, one can ask whether we can constrain any
possible non-commutativity in four dimensions from present
experiments.  The most striking feature of such non-commutativity
is likely to be the violation of Lorentz invariance.

The authors of \cite{harveyetal}, for example, have investigated the
violations of Lorentz invariance which would arise in a
non-commutative version of QED.  To proceed, these authors take
the tree level lagrangian of the theory, and perform the
Seiberg-Witten map to rewrite the theory in terms of conventional
quantum fields.  This generates a number of Lorentz-invariance
violating operators, all with two derivatives relative to the renormalizable
terms, i.e. dimension six.  This is basically because the indices on 
$\theta_{\mu \nu}$
must be contracted with derivatives.  They argue that
these constrain the parameter $\theta$ to be of order $(1 -
10^{-2}) {\rm TeV}^{-2}$.  Far stronger bounds arise if one considers 
potential
non-commutative effects in strong interactions\cite{stronglimits}.

On the other hand, one can imagine that operators of lower
dimension, such as \beq {\cal O}_1 = m_e\theta^{\mu \nu}\bar \psi
\sigma_{\mu \nu} \psi ~~~~~
{\cal O}_2 = \theta^{\mu \nu}
\bar \psi D_{\mu} \gamma_{\nu} \psi
~~~~~{\cal O}_3 = (\theta^2)^{\mu \nu}
F_{\mu \rho} F_{\nu}^{\sigma}~~~{\cal O}_4 = \theta^{\mu \nu}
\theta^{\rho \sigma} F_{\mu \nu} F_{\rho \sigma} \eeq
might appear at some level.
Indeed, we will demonstrate in this paper that these operators are
generated by quantum effects, in a variety of non-commutative
theories. Typically they are generated at two loop order. Because these
operators are (effectively) dimension four, while $\theta$ has dimensions of
inverse mass-squared, the coefficient must involve additional dimensionful
factors.  As we will see, if we suppose that the theory has some cutoff,
$\Lambda$, then there are two interesting regimes.  In the first, $\theta \Lambda^2
\gg 1$, and the additional dimensions are just made up by factors of
${\rm Tr}(\theta^2)$.  In other words, {\it the coefficients are independent
of the magnitude of $\theta$!}  In this case, such non-commutativity is
ruled out, no matter what the scale at which it might be relevant.  This is
an example of the infrared-ultraviolet connection.

On the other hand, such a cutoff is probably not sensible for other
reasons, connected with the infrared-ultraviolet connection.  For example,
scalar propagators have bizarre infrared singularities\cite{mrs}.
In the limit $\theta \Lambda^2 \ll 1$, the coefficients of these
operators are proportional to this dimensionless quantity.  The
limits on such  operators are quite stringent.  For example,
${\cal O}_1$ has the structure of a coupling of the electron
magnetic moment to a background magnetic field (though there is no
corresponding orbital coupling) . Just given that one can detect
the earth's magnetic field with ordinary magnets, one should be
able to establish a limit of order $10^{-15}$ on its dimensionless
coefficient. Indeed, the actual limits are orders of magnitude
more stringent, particularly from clock comparison
tests\cite{kosteleckya}
and from spin-polarized solids \cite{kosteleckyb}.
${\cal O}_2$ would lead to
propagation of photons with different speeds depending on their
polarizations. Again, stringent limits exist, here from cosmic
birefringence\cite{caroll2}.
However, because the coefficient is quadratic in $\theta$, the
corresponding limits on $\theta \Lambda^2$ are significantly
weaker.

In any case, even in this limit,  if we combine the limit on $\theta \Lambda$
with plausible restrictions on $\Lambda$, we can set very strong
limits on $\theta$.  For example, if $\Lambda = 1~{\rm TeV}$, than
calling $\theta = 1/M^2$, we will see $M>10^{12-13} GeV$, or 
possibly stronger.  
The limit
scales linearly with $\theta$.

\section{Operators Involving Photons}

We consider first operators of the type ${\cal O}_3$.  These are
generated already at one loop, in a non-commutative version of
QED.  Related computations have been done in \cite{susskindetal}.
The required Feynman rules can be found there, and in
\cite{ncqed}.  A straightforward computation gives, for the vacuum
polarization:
\beq
i\Pi_{\mu \nu}(q)  = -4e^2 \int {d^4 k \over (2 \pi)^4}\sin^2(k
\wedge q)
[4 k_{\mu} k_{\nu} +(k^2 g_{\mu \nu} ~{\rm pieces})]{1 \over k^4}
\eeq
where, noting that we will be interested in the leading terms at
small momentum, we have suppressed the $q$ dependence except that
from the Moyal factors.
If we suppose that the cutoff in the theory is much larger than
$1/\theta$, then this expression, while ultraviolet finite, is
singular at small momenta.  Instead, we can consider the limit
$\theta \Lambda^2 \ll 1$.   Now we have to ask precisely how we
cut off the theory in a gauge-invariant fashion.  We will not
investigate this question carefully here\footnote{One approach to
a cutoff starts with the observation of \cite{susskindetal} that
in a supersymmetric theory, this contribution is canceled by the
contribution of gauginos.  If we introduce a soft mass for the
gauginos, then, this could act as a regulator.}.  Instead, we note
that the operator ${\cal O}_4$ receives a contribution from the
first term in the integrand, and this contains
terms that are at least formally gauge
invariant.  Introducing a momentum-space cutoff yields:
\beq
{\cal L}_{eff} \approx
-{e^2 \over 16 \pi^2} \Lambda^4 {\cal O}_4.
\eeq

So already at one loop, Lorentz-invariant dimension four terms are
present.  As we will see, the experimental limits on such terms
are impressive.  But limits on the operator ${\cal O}_1$ are
potentially much stronger, given that it depends linearly on
$\theta$.  In the next section, we study this operator.

\section{Two loop Contributions to the Fermion Lagrangian}

We consider a Yukawa theory, with lagrangian,
\beq
{\cal L} = i \bar \psi \dslash \psi +  {1 \over 2}
(\partial \phi)^2
 + g \bar \psi \star \phi \star \psi - m \bar \psi \psi 
\eeq
It is important here that $\psi$ is a Dirac fermion; otherwise the
coefficient of 
the would-be operator, ${\cal O}_1$, vanishes 
\footnote{This is 
because the phase factor appearing at 
the vertex is a cosine for a Majorana fermion, but an 
exponential for a Dirac fermion.}.
The Feynman rules for
this theory are given, for example, in \cite{mrs}.  To study the operator
${\cal O}_1$, we study the fermion self energy evaluated on--shell.  
At one loop, the
only diagram is planar, and there is no $\theta$-dependence.  At
two loops, however, there is a non-planar diagram.
This diagram has non-trivial $\theta$-dependence.  In
the limit that $\theta \Lambda^2 \ll 1$, we can expand the integrand
in powers of $\theta$.  There is a term proportional to
$\sigma_{\mu \nu}$ which is quadratically divergent:\footnote{Here
and below, we do not carefully distinguish the operator ${\cal
O}_1$ from operators such as $\theta^{\mu \nu} \bar \psi
\sigma_{\mu \nu} \Dslash \psi$, which are equivalent if one uses
the equations of motion.  We will confine our attention here to
on-shell computations.}
\beq
4 m  \int {d^4k \over (2 \pi)^4} {d^4 \ell \over (2 \pi)^4}
\ell \wedge k \kslash \lslash {1 \over
k^4 (k+ \ell)^4 (\ell)^2} ~,
\label{wedgeeqn}
\eeq
where $ \ell \wedge k \equiv \ell_{\mu} \theta^{\mu \nu} k_{\nu}$.
Simplifying, this yields:
\beq
4 m i\int {d^4k \over (2 \pi)^4} {d^4 \ell \over (2 \pi)^4} k^{\mu}
k^{\nu} \ell^{\rho} \ell^{\sigma} \theta_{\mu \rho} \sigma_{\nu
\sigma} {1 \over
k^4 (k+ \ell)^4 (\ell)^2}
\eeq
To perform the integral, it turns out to be simplest to first combine the first
two terms using a Feynman parameter.  This yields
\beq
24 i m \int {d^4 k d^4  \ell \over (2 \pi)^8} d x {x(1-x) k \wedge \ell k_{\mu}
\ell_{\nu} \sigma^{\mu \nu} \over [k^2 + \ell^2 x(1-x)]^4 \ell^2}
\eeq
One can now do the integral over $k$, leaving
\beq
{m \over 2 \pi^2}{1 \over  (16 \pi^2)^2} {\theta^{\mu \nu} \sigma_{\mu \nu}
\over \ell^2}
\eeq
so, introducing a simple momentum space cutoff $\Lambda^2$ on $\ell^2$,
we obtain
\beq
{\cal L}_{eff} =
{1 \over 2 } m \Lambda^2 \left({g^2 \over 16 \pi^2}\right)^2
\theta^{\mu \nu} \overline{\psi} \sigma_{\mu \nu} \psi
~.
\eeq
The main point here is that the result is non-zero, and, up to
factors of order one, is of the size one might naively guess.

In the limit $\theta \Lambda^2 \gg 1$, one can also easily obtain
the leading piece of the integral.  One now must keep the full
exponential factor, $e^{i k \wedge \ell}$.  However, in some ways, this
is simpler.  We now have to study
\beq
\int d^4k d^4 \ell {4 m k_{\mu} \ell_{\nu} \sigma^{\mu \nu} e^{i k
\wedge \ell} \over k^4 (k + \ell)^4 \ell^2} ~.
\label{infL}
\eeq
Although by simple power counting 
this integral is both ultra--violet and infra--red 
divergent, the dependence on $\theta_{\mu \nu}$ is 
finite. To see this, first regulate the  
infra--red divergence by temporarily inserting an 
infra---red cutoff. For low momentum we may then 
expand  
the exponential 
to obtain a power series in $\theta$. The 
leading term is independent of $\theta$ and 
is divergent. But the 
integral for the subsequent terms depending 
on $\theta$ are convergent due 
to the additional powers of the momenta. For these terms 
we may remove the cutoff.  
In the ultra--violet the phase factor damps the 
logarithmic 
divergence, and again the $\theta$ dependence is finite. 
  
This integral also has the 
interesting feature that it is independent of 
the overall scale of $\theta_{\mu \nu}$. To see this, 
write $\theta _{\mu \nu} = \theta b_{\mu \nu}$ 
where $b_{\mu \nu}$ is a matrix of numbers describing 
the orientation of $\theta_{\mu \nu}$ relative to 
a fixed coordinate system. By rescaling the 
momenta appearing in the integral the dependence 
on $\theta$ can be eliminated. Remarkably, our result 
in the limit of a large cutoff 
depends only on the direction of 
$\theta_{\mu \nu}$ and not its magnitude.  

Now in the integral, we can proceed much as we did before, and
obtain an analytic result in a few lines of algebra.  
First simplify the integral by 
setting the non-commutativity to lie, say, in
the $1,2$ direction.
Then 
introduce a Feynman parameter to combine the first two factors in
the denominator.  One then shifts $k$ in the usual way, and
obtains
\beq
24 m \int {dx d^4 k d^4
\ell x (1-x) \sigma_{\mu \nu} k^{\mu} l ^{\nu} 
 \over [k^2 + \ell^2 x(1-x)]^4 \ell^2} e^{i
\theta_{12} (k_1 \ell_2 - \ell_1 k_2)} ~.
\label{infL2}
\eeq
Despite the shift the
numerator retains the same form as before 
because $\sigma_{\mu \nu}$ and $\theta_{\mu \nu}$ 
are anti--symmetric. The integral involving 
components of $\sigma_{\mu \nu}$ 
other than $\sigma_{12}$ now vanish. This is because only 
for the 1,2 components is the integrand {\em not} an 
odd function of the momenta. Then (\ref{infL2}) simplifies 
to 
\beq
24 m \sigma_{12} {1 \over i} {d \over d \theta} \int {dx d^4 k d^4
\ell x (1-x) \over [k^2 + \ell^2 x(1-x)]^4 \ell^2} e^{i
\theta_{12} (k_1 \ell_2 - \ell_1 k_2)} ~.
\eeq
Now one can rescale $\ell$ to eliminate the $x(x-1)$ in the
denominator, and combine the remaining two factors with a new
Feynman parameter.  One can then integrate over the components of $k$
and $\ell$ in the $0$ and $3$ directions.  Introducing a Schwinger
parameter to exponentiate the remaining denominator, one can
sequentially do the integrals over the $1$ and $2$ components of
$k$ and $\ell$.  All of the integrals involved are elementary, and
one obtains, finally:
\beq
{\cal L}_{eff}
= {4 \over 3}m \left ({g^2 \over 16 \pi^2}\right )^2 {\pi
\theta^{\mu \nu}\over \sqrt{{1 \over 2}{\rm Tr}\theta^2}}
\overline{\psi} \sigma_{\mu \nu} \psi ~.
\eeq

These results are readily extended to $U(1)$ and $U(N)$ gauge
theories containing fundamental 
matter. 
For the case of a $U(1)$ theory, 
there are two diagrams
at two loop order \footnote{Two-loop diagrams with 
one-loop self-energy subgraphs do not 
contribute to ${\cal O}_1$.}. The Feynman rules 
may be found in \cite{ncqed}, after correcting a 
sign in the phase of the photon--electron--electron Feynman rule 
so that the vertex factor is proportional to 
$e^{i p_I \wedge p_F}$, for 
incoming (outgoing) electron momentum $p_I$ $(p_F)$ and 
using the $\wedge$ notation defined below (\ref{wedgeeqn}) .  
One finds that these two diagrams contribute equally but 
add constructively, 
in both limits.
In the case of a U(N) gauge theory,
there are again two diagrams.  The Feynman rules are given, for example,
in
\cite{armoni} \footnote{In this reference the momenta are
pointing into the vertices. This is confirmed by taking the 
commutative limit to obtain 
the usual rule for non-abelian gauge theory.}. 
Here one also does not find a cancellation 
\footnote{In 
an earlier version a cancellation for both the 
$U(1)$ and $U(N)$ theories was obtained. This result was 
incorrect due to a sign error in deriving a Feynman rule.}. 
In fact the
result is independent of $N$, so the $U(N)$ and $U(1)$ theories 
generate the operator with the same coefficient.  
One finds 
in the limit $\theta \Lambda ^2 \ll 1$ the effective Lagrangian 
for the on-shell amplitude is 
\beq
{\cal L}_{eff} = {3 \over 4} 
 m \Lambda^2 \left ({e^2 \over 16 \pi^2}\right )^2  \theta^{\mu \nu}
\bar{\psi} 
\sigma_{\mu \nu} \psi ~,
\eeq
where the $U(1)$ generator is normalized to $1/2$. 
In the limit $\theta \Lambda^2 \gg 1$ it is given 
by
\beq
{\cal L}_{eff} = 2 m  \left ({e^2 \over 16 \pi^2}\right )^2
{\pi  
\theta^{\mu \nu}\over \sqrt{{1 \over 2}{\rm Tr}\theta^2}} 
\bar{\psi} \sigma_{\mu\nu} \psi ~.
\eeq

Finally, we can consider a theory with Yukawa interactions and a
U(1) gauge interaction.
In this case, there is again a non-zero
contribution.  Proceeding as above, we obtain:
\beq
{\cal L}_{eff} =- m \Lambda^2
 \left({e g \over 16 \pi^2}\right)^2 \theta^{\mu \nu}
\overline{\psi} \sigma_{\mu \nu} \psi
\eeq
where $e$ and $g$ denote the gauge and Yukawa couplings,
respectively. In a U(N) gauge
theory 
this is generalized to
\beq
{\cal L}_{eff} = -{N \over 2 } m \Lambda^2
 \left({e g \over 16 \pi^2}\right)^2 \theta^{\mu \nu}
\overline{\psi} \sigma_{\mu \nu} \psi ~.
\eeq
We will discuss the possible experimental implications of these results
below.

These calculations have all been done in terms of the
non-commutative variables.  Because we have calculated operators
which are bilinear in the fields, the non-commutativity is not
relevant.  We have considered how these computations might look if
one first performs the Seiberg-Witten map.  Indeed, the
calculations are distinctly more complicated.  At one loop, it is
obvious that there is no $\theta$-dependence in the self energy,
if one works with the non-commutative variables.  If, in a $U(1)$
theory, say, one first performs the Seiberg-Witten map, there are
many individual diagrams with non-trivial $\theta$ dependence.  Off shell,
working, say, at zero momentum, it is easy to see that there are terms in the
effective action with non-trivial $\theta$-dependence.  We
have checked that the $\theta$-dependence vanishes on shell.  This
seems quite reasonable.  Thinking of the Seiberg-Witten
transformation as a field redefinition, we do not expect
correlation functions of the new fields to be the same as those of
the original ones, but we do expect on-shell quantities to be the
same.

In general, then, we see that the operator ${\cal O}_1$ is
generated. In the small $\Lambda \theta$ limit, we expect the
experimental bounds to be very strong.  We will discuss these
bounds in the next section.  One might object that the cutoff we
have introduced is artificial.  A natural cutoff for
noncommutative field theory is to realize it as a large $N$ limit
of lower dimensional field theories of $N\times N$ matrices
(though the precise nature of the limit has not yet been
understood). In this way one appears to guarantee that
counterterms will all involve $*$-products. However, precisely in
the two loop nonplanar graphs that we have studied, one encounters
graphs that produce double trace operators.   These are the
diagrams that exhibit UV-IR mixing reminiscent of the exchange of
closed string states in open string field theory.  It is important
then, that we have done the two loop calculation in the infinite
cutoff limit as well, and found a relevant operator with a
coefficient independent of the scale of
$\theta$.   One can easily imagine that UV-IR
mixing will produce all sorts of bizarre effects that might lead
to experimental signatures.  This has not been studied.   Our
results show that in addition to these peculiar effects, the NC
field theory produces a relevant Lorentz violating term with large
coefficient.   The term is precisely of the form we found with an
ad-hoc cutoff much lower than the energy scale of noncommutivity.
Thus, the constraints on a noncommutative field theory with no
cutoff are much more stringent than those we will exhibit in the
next section.

\section{Experimental limits}

One expects that there are extraordinary limits on the coefficient
of the operator ${\cal O}_1$.  For example, from the fact that one
can measure the earth's magnetic field with a compass, one can set
a limit on the coefficient of ${\cal O}_1$ of order $10^{-15}~{\rm
MeV}$.  From precision measurements in atomic systems, one expects
to be able to set very strong limits.  In fact, the best limit on
this operator for the electron comes from
magnetic
systems\cite{kosteleckyb}, where a limit of approximately
$10^{-25} ~{\rm MeV}$ is set \cite{uwash2}.
This bound is obtained from studying the oscillation
of a highly electron spin-polarized
torsion pendulum, where the presence of the
operator ${\cal O}_1$ and the rotation of the earth
induces a time-dependent
macroscopic
torque
on the pendulum \cite{uwash1,uwash2}.
The limit from precision
tests of hyperfine splitting
is about one order of magnitude weaker \cite{kosteleckya}.

A number of authors have studied limits on dimension six operators
proportional to ${\theta}$.  In non-commutative QED, one can set limits on
$\sqrt{\theta}$
in the several TeV range\cite{harveyetal}.
The strongest limit is that discussed
in \cite{stronglimits}.  These authors assume that
$\theta$-dependent terms in a non-commutative version of QCD lead
to a coupling $\theta^{\mu \nu} \bar N \sigma_{\mu \nu} N$, where
$N$ is the nucleon wave function.  This leads to a striking limit
on $\theta$, $\sqrt{\theta} < (10^{15} {\rm GeV})^{-1}$.

If we attempt to take the cutoff to infinity, it is clear that we
can rule out non-commutativity at any scale.  
In this limit the coefficients of 
the Lorentz--violating operators 
are far too large to accommodate the 
experimental bounds, for they are 
suppressed only by loop and gauge or Yukawa factors and 
are independent of the size of the non--commutative scale.
This is a reflection
of the infrared--ultraviolet connection.  On the other hand,
as in \cite{hinchliffe}, it is more reasonable to include an explicit
cutoff.  Otherwise, one will obtain, for example, unacceptable infrared
behavior for gauge boson and scalar propagators.

Lacking a complete
non-commutative generalization of the standard model, it is hard to set
precise limits.  Still, we have seen that in theories with Yukawa and
gauge couplings, there are contributions to ${\cal O}_1$ which proportional to
$\alpha y^2$, where $y$ is the Yukawa coupling, and contributions 
proportional to $\alpha^2$. The latter contributions 
are identical in both the pure gauge $U(N)$ and 
$U(1)$ theories. This suggests the possibility, although 
unlikely, that in 
a non--commutative formulation of $SU(N)$
the 
two--loop pure gauge contribution could vanish. To verify or 
disprove this speculation would require a computation 
in a satisfactory formulation of $SU(N)$ non--commutative 
theories. 
Given these remarks, we
make the conservative choice of obtaining 
bounds from only the pure $U(1)_{em}$ and mixed gauge--Yukawa
contributions and ignore the much stronger 
bounds obtained from the pure $U(3)_c$ contribution.
 

 
For the electron, in particular, the contributions proportional 
to $\alpha_{em} \lambda^2$ 
translate to a limit on the
dimensionless quantity $\theta \Lambda^2$, roughly:
\beq
\theta \Lambda^2 < 10^{-8}.
\eeq
There are also much larger 
contributions proportional to $\alpha^2_{em}$. 
For the electron, these translate to a limit on the
dimensionless quantity $\theta \Lambda^2$, roughly:
\beq
\theta \Lambda^2 < 10^{-19} ~.
\eeq
Given that the operator is suppressed by a power
of the electron mass, it might be advantageous to study Lorentz
violating constraints in muons. A reanalysis of Zeeman
hyperfine splitting data in muonium, looking for sideral
time variations, could
constrain the coefficient of ${\cal O}_1$ to be less than
$10^{-19}$ MeV \cite{bluhmmuon}.
Future data from the BNL $g-2$ experiment
could improve the latter limit
by two to three orders of magnitude
\cite{bluhmmuon}. Putting in the numbers, the latter
limit would only 
constrain $\theta \Lambda^2$ at the same level
as the torsion pendulum experiment, whereas the
hyperfine data provide a weaker bound.

From the neutron and proton, we obtain a stronger limit.   The
precise value requires translating the quark moments to nuclear
moments, and we will content ourselves with very crude estimates.
The strongest experimental limit comes from the
neutron\cite{kosteleckya}, where the limit is $10^{-27} {\rm
MeV}$. First consider the contributions proportional 
to the Yukawa couplings and the strong coupling.   
For the up and down quarks,
this translates, roughly, to a limit on the dimensionless
combination
\beq \theta \Lambda^2 < 10^{-17} ~.
\eeq
We can do better using the strange quark. Here however
we need to know what fraction of the nucleon spin is due
to the strange quark, for which there is both considerable
theoretical and experimental uncertainty. The SAMPLE
collaboration has recently measured the strange contribution
to the nucleon magnetic moment,
and find $\mu_s = 0.01 \pm 0.29 \pm 0.31 \mu_N$ \cite{sample},
where $\mu_N$ is the Bohr magneton.
Using the central value
a rough limit is
\beq
\theta \Lambda^2 < 10^{-19} ~.
\eeq

Let us now consider the pure electromagnetic 
contributions. 
It is reasonable 
to expect that the matrix element of this contribution 
does not vanish for either the proton 
or neutron. Here though the 
experimental limit is weaker for the proton, with a bound of roughly 
$10^{-24}$ MeV \cite{kosteleckya}. 
This translates into a limit 
\beq 
\theta \Lambda^2 < ~ 10^{-19},
\eeq 
which is comparable to the bound obtained from the electron. 
The limit from the neutron is  
stronger, roughly : 
\beq 
\theta \Lambda^2 < ~ 10^{-22},
\eeq 

A three--loop pure $SU(3)_c$ contribution provides a 
stronger limit. 
For the neutron we
would obtain $\theta \Lambda^2 < 10^{-23}$.
Limits on ${\cal O}_2$ involving the nucleon also
exist, but are typically three orders of magnitude weaker
than ${\cal O}_1$ \cite{kosteleckya}. A non-vanishing pure gauge
contribution at two-loops, could be competitive with the
limits found here. Work is in progress in evaluating
the contribution to this operator \cite{alexey}.

We can also consider
Lorentz violating operators involving the photon. Starting at two
loops, for example, the following dimension 4 operators
may be generated,
\beq
{\cal O}_{3}
= \lambda_3 (\theta ^2)_{\mu \nu} (F^2)^{\mu \nu} ~~~,
{\cal O}_4 = {\lambda_4 \over 8}
(\theta_{\mu \nu} F^{\mu \nu})^2 ~~~,
~~~ {\cal O}_{5}
= \lambda_5
\theta_{\mu \rho} F^{\rho \sigma} \theta _{\sigma \tau } F^{\tau \mu}
~.
\label{op}
\eeq
The $\lambda_i \sim  (\alpha/4 \pi )^2 \Lambda ^4$ 
and their detailed values are unimportant for this discussion.   
These Lorentz--violating operators 
introduce $B_i B_j$ and
$E_i E_j$ terms into the action, and they affect the propagation
of light in vacuum, causing the vacuum to behave much like an
anisotropic
dielectric medium \cite{ck2}. In particular, light
travels at different speeds depending on the directions of
polarization and propagation. This leads to cosmic birefringence,
for which stringent limits already exist \cite{caroll2,ck2}.

To understand this, for simplicity
consider ${\cal O}_4$ only. The behavior discussed
here will also apply to the other operators.
Assume that $\theta$ is non-zero
in the 1-2 direction. Solving the field equations
one finds that for light moving only in the 3-direction
$p^0 = |p_3|$. Similarly, light
moving in either the 1 or 2 directions but polarized along the
3 direction is unaffected. But light moving along the 1-direction say,
and polarized in the 2-direction has a modified dispersion relation,
$p^0 =(1+ \lambda_2 \theta^2/2+\cdots)|p_1|$.  The modification
for more general polarizations and directions of propagation is
easily worked out.

The experimental limit on this effect is
obtained from studying polarized
light from distant radio sources, and looking
for a dependence of the angle of polarization on the
distance to the source \cite{caroll2}.
More specifically, \cite{radio} studied a large number
of radio sources, and found a strong correlation between
the angle of polarized light and the major axis of
the source (after correcting for Faraday rotation).
For the active radio sources this angle is roughly
$90^0$, indicating that the magnetic fields of these
sources runs parallel to the major axis, and that the light is
produced by synchrotron
radiation.
If Lorentz violation is present, then the angle of
the polarized light relative to the major axis
will be rotated away from
$90^0$ by an amount that grows with the distance to
the source.
Examining this data, Carroll, Field
and Jackiw do not find any redshift dependence \cite{caroll2}.
This constrains the
relative phase given above
to be $\phi = \delta \vec{p} \cdot \vec{x}
\ltap {\cal O}(1)$. Since
the galaxies observed in \cite{radio}
span a sizable fraction of
the observable Universe,
the constraint is
\beq
{\vert \delta \vec{p} \vert
 \over H} \ltap {\cal O}(1) \Rightarrow \delta p \ltap 10^{-42}
~\hbox{GeV}  ~,
\label{pbound}
\eeq
with the limit quoted in \cite{caroll2} smaller by $h_0 /4$ .
Strictly speaking, \cite{caroll2} studied the effect of
a different set of operators,
namely those of the type $E_i B_j$. These lead to a dispersion relation
that is independent of wavelength \cite{caroll2,ck2},
for which they obtain
the bound quoted above. As \cite{ck2} argue,
we cannot directly translate this limit to a
bound on the operators in (\ref{op}) since the dispersion relation is
no longer independent of wavelength, making it is more difficult
to disentangle the Lorentz violating effect from Faraday
rotation.
In principle, however, these effects could
be distinguished since the Faraday
rotation depends on the square of the wavelength, whereas
for the Lorentz violating effect the phase scales inversely with
wavelength.
Following \cite{ck2},
it is then reasonable that the data should
still imply a bound that is
roughly given by (\ref{pbound}).
As the wavelength of the radio sources analyzed in \cite{radio}
are typically 10 cm, inserting this value in the previous
dispersion relation leads to the limit of
\beq
\lambda_i \theta^2 < 10^{-28} ~.
\eeq
This is an impressive limit, but because
$\theta$ appears
quadratically,
it is not competitive with
the terrestrial constraints. For inserting the
loop factor suppression in $\lambda _i \sim 
(\alpha/ 4 \pi)^2 \Lambda^4$ leads to the bound
\beq
\theta \Lambda^2 < 10^{-12} ~.
\eeq

Of course, without a detailed underlying theory, these limits
cannot be interpreted unambiguously.  With plausible assumptions
about the cutoff, we can bound $\theta$ at an extraordinary
level.  Even if the cutoff is $1 ~{\rm TeV}$, we can
set a limit on $\theta$, $\theta < (10^{12-13} $GeV$)^{-2}$.

\noindent

{\bf Acknowledgements:}

\noindent

We thank A. Armoni, Jeff Harvey, Nemanja Kaloper 
and Scott Thomas,
for conversations and comments. We also thank 
Mahiko Suzuki, who has similar 
observations, for stimulating discussions. 
This work supported in part by the U.S.
Department of Energy.



\begin{thebibliography}{99}

\singlespaced

\bibitem{ncreview}
For a review with extensive references, see M.R. Douglas and N.A.
Nekrasov, ``Non-Commutative Field Theory," hep-th/0106048, to
appear in Reviews of Modern Physics.

\bibitem{sw}
N. Seiberg and E. Witten, ``String Theory and Non-Commutative
Geometry," hep-th/9908142, JHEP {\bf 09} (1999) 032.

\bibitem{rizzo}``Signals for Noncommutative Interactions 
at Linear Colliders'', J. Hewett, F. Petriello, and 
T. Rizzo,  hep-ph/001035
Phys.Rev.{\bf D64} :075012,2001. 

\bibitem{stronglimits}
I. Mocioiu, M. Pospelov and R. Roiban,
``Low Energy Limits on the Antisymmetric
Tensor Field Background on the Brane and the Non-Commutative
Scales,"
Phys. Lett. {\bf B489}
(2000) 390,  hep-ph/0005191.

\bibitem{ncqed}
I. Riad and M.M. Sheikh-Jabbari, ``Noncommutative QED and
Anomalous Dipole Moments," hep-th/0008132.

\bibitem{susskindetal}A. Matusis, L. Sussking, and
N. Toumbas, ``The IR/UV Connection in Noncommutative
Gauge Theories'', JHEP {\bf 0012} (2000) 002,
hep-th/0002075.

\bibitem{hinchliffe}
I. Hinchliffe and N. Kersting, ``CP-violation
from Noncommutative Geometry,"  hep-ph/0104137.

\bibitem{harveyetal}
S. Carroll, J.A. Harvey, V. Alan Kostelecky, C.D. Lane and T.
Okamoto, ``Non Commutative Field Theory and Lorentz Violation,"
hep-th/0105082.

\bibitem{jackiw}
Z. Guralnik, R. Jackiw, S.Y. Pi and A.P. Polychronakos, ``Testing
Non-Commutative QED, Constructing Non-commutative MHD",
hep-th/0106044.

\bibitem{mrs}
S. Minwalla, M. Van Raamsdonk and N. Seiberg, ``Non-Commutative
Perturbative Dynamics," JHEP {\bf 0002} (2000) 20, hep-th/9912072.

\bibitem{kosteleckyb}
R. Bluhm and V. Alan Kostelecky, ``Lorentz and CPT tests with
Spin-Polarized Solids," Phys.Rev.Lett.{\bf 84} (2000) 1381, hep-ph/9912542.

\bibitem{uwash2}M.G. Harris, Ph.D thesis, University of
Washington, 1998.

\bibitem{uwash1}E.G. Adelberger {\em et al.}, in P. Herczeg {\it et al.},
eds., {\em Physics Beyond the Standard Model}, p. 717, World
Scientific, Singapore, 1999.

\bibitem{kosteleckya}
V.A. Kostelecky and C.D. Lane,
``Constraints on Lorentz Violation
from Clock-Comparison Experiments,"
Phys.Rev. {\bf D60}  (1999) 116010, hep-ph/9908504.

\bibitem{hayakawa}M. Hayakawa, ``Perturbative analysis 
on infrared and ultraviolet aspects of noncommutative QED 
on $R^4$'', hep-th/9912167. 

\bibitem{armoni}A. Armoni, 
``Comments on Perturbative Dynamics of Noncommutative Yang-Mills 
Theory'', 
Nucl.Phys. {\bf B593} (2001) 229, hep-th/0005208.

\bibitem{bluhmmuon} R. Bluhm,
V.A. Kostelecky, and C. Lane, ``CPT and Lorentz Tests with Muons'',
Phys. Rev. Lett. {\bf 84} (2000) 1098, hep-ph/9912451


\bibitem{sample}SAMPLE Collaboration (R. Hasty {\em et. al.},
``Strange Magnetism and the Anapole Structure of the Proton'',
Science {\bf 290} (2000) 2117, nucl-ex/0102001.

\bibitem{alexey}A. Anisimov, in progress.

\bibitem{ck2}D. Colladay and V.A. Kostelecky, ``Lorentz--Violating
Extension of the Standard Model'', Phys. Rev. {\bf D58 } (1998)
116002,
hep-ph/9809521.

\bibitem{caroll2}S. Carroll, G. Field and R. Jackiw,
``Limits on a Lorentz and Parity Violation Modification
of Electrodynamics'', Phys. Rev. {\bf D41} (1990) 1231.

\bibitem{radio}P. Haves and R. G. Conway, Mon. Not. R. Astr. Soc.
{\bf 173} (1975) 53; J. N. Clarke, P.P. Kronberg and M.
Simard-Normandin, {\em ibid.}, {\bf 190} (1980) 205.

\end{thebibliography}
\end{document}